\documentclass[prb,twocolumn,amssymb,showpacs]{revtex4}
\usepackage{graphicx}
\usepackage{amsmath}

\begin{document}

\title{Evolution of the magnetic phase transition in MnO confined to channel type matrices.
Neutron diffraction study.}
\author{I. V. Golosovsky}
\affiliation{Petersburg Nuclear Physics Institute, 188300, Gatchina, St. Petersburg, Russia.}
\author{I. Mirebeau}
\affiliation{Laboratoire L\'eon Brillouin, CE-Saclay, F-91191, Gif-sur-Yvette, France.}
\author{V. P. Sakhnenko}
\affiliation{Rostov State University, 344090, Rostov/Don, Russia.}
\author{D. A. Kurdyukov and Y. A. Kumzerov}
\affiliation{A. F. Ioffe Physico-Technical Institute, 194021, St. Petersburg, Russia.}

\begin{abstract}
Neutron diffraction studies of antiferromagnetic MnO confined to MCM-41 type matrices with channel diameters
$24 {\div} 87$ {\AA} demonstrate a continuous magnetic phase transition in contrast to a discontinuous first
order transition in the bulk. The character of the magnetic transition transforms with decreasing channel
diameter, showing the decreasing critical exponent and transition temperature, however the latter turns out
to be above the N\'eel temperature for the bulk. This enhancement is explained within the framework of Landau
theory taking into consideration the ternary interaction of the magnetic and associated structural order
parameters.
\end{abstract}

\pacs{61.12.Ld; 74.78.Na; 75.30.Kz} \maketitle

\section{Introduction}

The properties of magnetics confined to nanometer scale cavities drastically differ from those of the bulk
that stimulates an interest in the behavior of magnetic materials in the conditions of so-called
"restricted" or "confined" geometry. One of the most intriguing problems is the influence of the "confined
geometry" on the magnetic phase transition.

We have recently reported that antiferromagnetic MnO embedded in a vycor-glass type matrix with a random
network of interconnected pores exposes some remarkable differences when compared to the bulk \cite{MnO-PRL}.
It was found by neutron diffraction that a first order magnetic transition becomes a second order transition
with the N\'eel temperature 120.2(1) K slightly higher than 117.6(1) K for the bulk MnO\cite{Lines}. It was
demonstrated that confined MnO forms aggregates with the average diameter of 145(3) {\AA}. Below the magnetic
transition a long range magnetic ordering appears within a "core" with a smaller average diameter of about
100(3) {\AA}, that leads to a reduction of the volume averaged magnetic moment with respect to the bulk
value.

We continued these studies with MnO confined to the cavities with another topology, namely, to nanochannels
of mesoporous MCM-41 type matrices. From synchrotron diffraction experiments we found that MnO confined to
the large 47-87 {\AA} channels crystallizes in the form of thin (about 10 {\AA}) ribbon-like structures with
a width of about the channel diameter and a length of 180-260 {\AA}. In the matrices with the narrower
channels of 24 and 35 {\AA} diameters MnO crystallizes in the form of nanowires with diameters of about 20
{\AA} and a length of about 180-200 {\AA} \cite{LURE}. The morphology of the MnO nanoparticles thus varies
considerably depending on the matrix and can show different dimensionalities of the spin system.

Besides synchrotron diffraction experiments we had performed electron spin resonance (ESR) experiments, which
confirmed the presence of the surface disordered spins in confined MnO \cite{ESR}. However the ordered
magnetic moment cannot be studied by the ESR technique and other experimental techniques are needed.

In the present study neutron diffraction is used to investigate in a more systematic way the size
effect on magnetism in MnO confined to channel-type matrices.

\section{Experiment }

\subsection{Experimental details}

The experiments were performed with MnO embedded in the channel of MCM-41 matrices \cite{Grun} with 24 and 35
{\AA} channel diameters and SBA-15 matrices \cite {Zhao} with 47, 68 and 87 {\AA} channel diameters,
respectively. These matrices differ by the preparation techniques and both present an amorphous silica
(SiO$_{2}$) with a regular hexagonal array of parallel cylindrical nanochannels. The matrices in the form of
powders with a grain size $\sim $ 1-2 ${\mu}m$ were prepared in the Laboratoire de Chimie Physique,
Universit\'e Paris-Sud, France \cite{Morineau-2}. All samples were filled with MnO from a solution by the
"bath deposition" method developed in the Ioffe Physico-Technical Institute (St Petersburg, Russia). The
consistent analysis of neutron diffraction, X-ray diffraction, ESR and magnetization measurements ensures
that MnO predominantly occupies the channel voids.

Neutron diffraction experiments were carried out at the diffractometer G6-1 of the Laboratoire L\'eon
Brillouin at the Orph\'ee reactor with a neutron wavelength of 4.732 {\AA}.

\subsection{Shape and dimension of magnetic domains}

\begin{figure} [t]
\includegraphics* [width=\columnwidth] {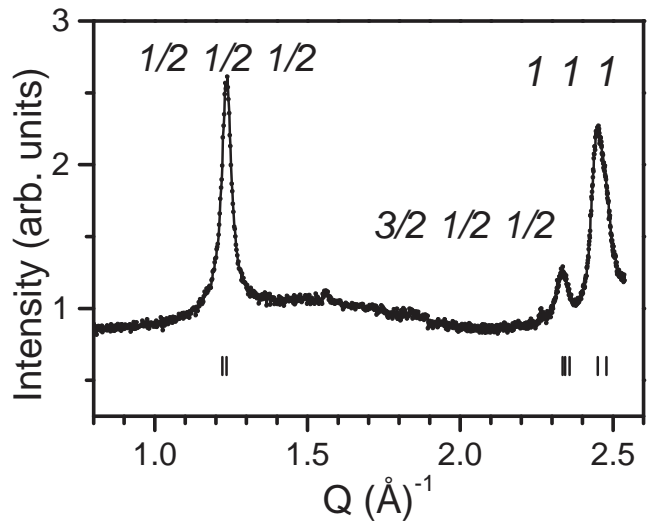}
\caption{Neutron diffraction patterns, observed (closed circles) and calculated (solid line) from MnO
confined within SBA matrix with 47 {\AA} channel diameter.} \label{profile}
\end{figure}

A typical neutron diffraction pattern measured at 10 K is shown in figure \ref{profile}. To increase
intensity we used a large neutron wavelength. However, in this case we could observe only two magnetic Bragg
reflections: $\frac{1}{2} \frac{1}{2} \frac{1}{2}$ and $\frac{3}{2} \frac{1}{2} \frac{1}{2}$ and one nuclear
reflection 111.

Synchrotron experiments demonstrated that the nuclear peaks from the samples with large channels (47-87
{\AA}) are asymmetric and have specific "saw-tooth" profiles, while the peaks from the samples with smaller
channels (24 and 35 {\AA}) are symmetric. By numerical calculation of the diffraction profiles it was shown
that the confined nanoparticles have a shape of nanoribbons and nanowires respectively \cite{LURE}.

However, the magnetic peaks appear to be symmetric for all samples. This suggests that in contrast with the
nuclear particles the magnetic domains should be considered as needle-shape objects whatever the channel
diameter.

\begin{figure} [t]
\includegraphics* [width=\columnwidth] {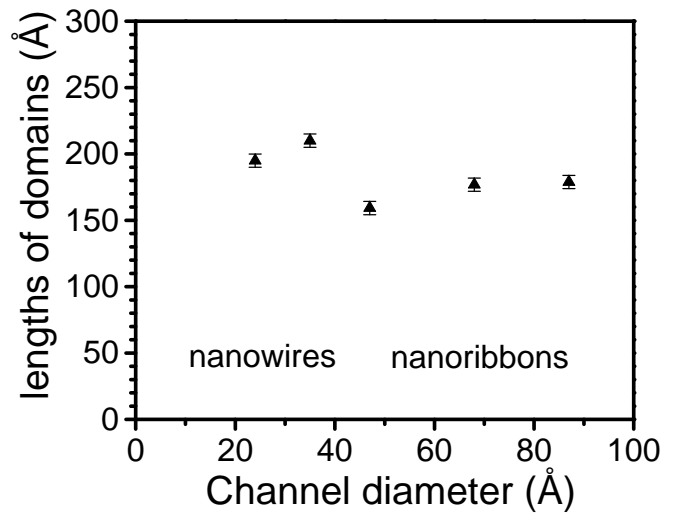}
\caption{The lengths of magnetic domains of MnO within the channels of different diameter.} \label{lengths}
\end{figure}

We evaluated the size of the magnetic domains from the peak broadening using the Thompson-Cox-Hastings
approximation of the lineshape \cite{Thompson}. It is well known that the smaller dimensions of the
diffracting objects contribute to the peak "pedestal", while the peak width is defined mainly by the largest
dimension. Therefore from the peak broadening we can reliably evaluate only the length of the magnetic
domains shown in figure \ref{lengths}. It is seen that the lengths of magnetic domains in the channels with
different diameter appear to be practically similar and equal to about of 180 {\AA}.

\subsection{Magnetic moments and phase transitions}

The indexing of the magnetic reflections corresponds to antiferromagnetic ordering of type-II in the fcc
lattice similar to the bulk MnO \cite{Shull}.

The volume averaged magnetic moment of confined MnO for all samples was calculated from the magnetic
reflection intensity. It appears noticeably smaller than the free-ion value of 5$\mu${$_{B}$} and the moment
in the bulk of 4.892 $\mu_B$/ion \cite{MomentBulk} and does not depend on the channel diameter in the limits
of our statistical accuracy.

The average measured value of 3.98(5) $\mu_B$/ion is close to 3.84(4) $\mu_B$/ion measured for MnO embedded
in a porous glass \cite{MnO-PRL} with the same magnetic structure. As well as in the last case the observed
moment reduction can be readily explained by the disorder of the magnetic moments at the surface.

\begin{widetext}
\begin{figure} [t]
\includegraphics* [width=6.5in] {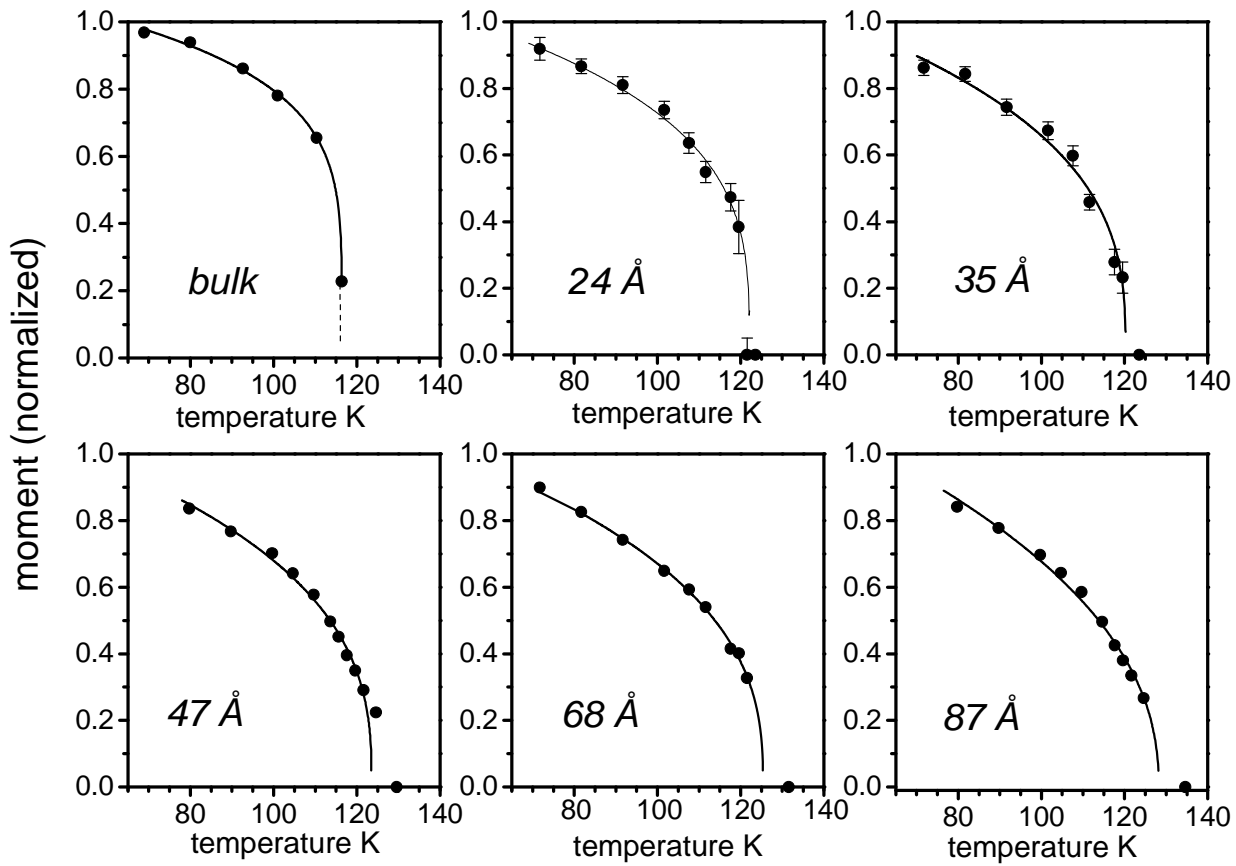}
\caption{Temperature dependence of the normalized magnetic moment for MnO confined to the channels of
different diameter. Solid line corresponds to a fit with a power law. Errors (estimated standard deviation),
if not shown, do not exceed the symbol size.} \label{magnetic}
\end{figure}
\end{widetext}

The temperature dependencies of the magnetic moment for nanoparticles embedded in the matrices with
different channel diameters are shown in figure \ref{magnetic}. For comparison the corresponding dependence
for the bulk \cite{Shull} is displayed too.

\newpage

\begin{figure} [t]
\includegraphics* [width=\columnwidth] {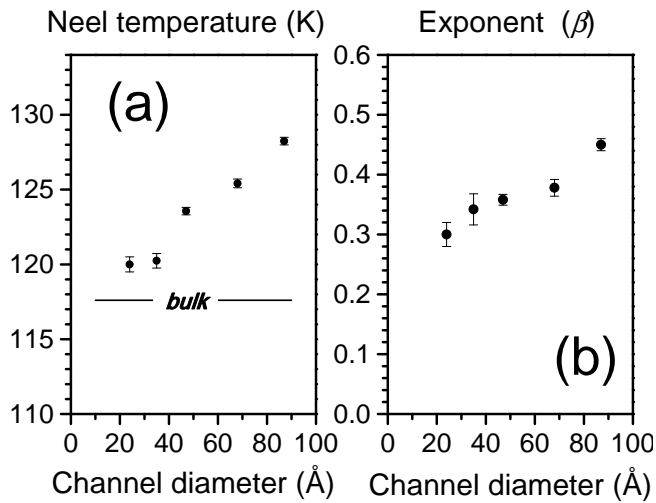}
\caption{ Dependencies of N\'eel temperature $T_N$ (a) and exponent $\beta$ (b) on channel diameters from the
fitting with a power law.} \label{exponent}
\end{figure}

It is seen that the magnetic transition in confinement becomes continuous with the enhanced temperature of
transition $T_N$ in contrast with the bulk ($T_N=118$ K). With decreasing the channel diameter the slope of
the temperature dependence increases, i.e. the character of the phase transition transforms. Fitting the
magnetic moment \emph{m} with a power law $m \sim (1-T/{T_N})^\beta$ demonstrates that the exponent $\beta$
and $T_N$ increase with increasing the channel diameter (Figure \ref{exponent}a and \ref{exponent}b).

\section{Discussion}

\subsection{Evolution of the magnetic phase transition}

As shown in the model of bilinear exchange \cite{Morosin} and by a group-theoretical method \cite{Mukamel}
the magnetic phase transition in the bulk MnO cannot be of a second order, and in fact, a discontinuous first
order transition is observed in experiment.

However the observation of the continuous phase transition in MnO confined within nanochannels is not
surprising and had already observed on MnO embedded in a porous glass \cite{MnO-PRL}. This finite-size
rounding is well understood and results from the correlation length being limited by the nanoparticle size
\cite{Imry,Challa}.

It is known that a magnetostriction interaction could affect the character of the magnetic transition. Indeed
in the bulk MnO the softening of the elastic modulus $c_{44}$, when approaching to the transition from above,
had experimentally observed\cite{Miyahara}. However, the values of the exponents $\beta$ (figure
\ref{exponent}a) and the continuous character of the magnetic transition (figure \ref{magnetic}) give
evidence that a fluctuation scenario dominates on our case. In the condition of the "restricted geometry" due
to the splitting of the multicomponent order parameter into a few order parameters of lower dimensionality,
there is no theoretical restriction for the character of the transition \cite{Brezin}.

In the bulk as well as in confined MnO the magnetic transition is accompanied by a rhombohedral crystal
distortion, which lifts the frustration in the first coordination sphere and stabilizes the antiferromagnetic
structure. This distortion yields a difference in the exchange interactions between spins within the
ferromagnetic layer and between spins in adjacent ferromagnetic layers. From neutron profile refinements the
crystal structure distortions for all studied samples appear to be the same, so it cannot be responsible for
observed transformation.

Since the lengths of magnetic domains in the channels of different diameter are similar we suggest that the
observed evolution of the phase transition should be attributed to decrease of the magnetic domain diameter
with channel diameter decreasing. It results in the increasing of anisotropy and in the changing of the
dimensionality of the magnetic system towards to quasi-one-dimensional case.

Surprisingly, the measured critical exponent is changing linearly with the channel diameter, i.e. with the
ratio of the volume/surface of nanoparticle. Supposing that at the surface the magnetic bonds are broken,
this ratio is proportional to the number of magnetic bonds in the magneto-ordered core. Therefore the
observed increasing the exponent $\beta$  with the channel diameter towards to the mean-field limit of 0.5
\cite{mean-field} (figure \ref{exponent}b) obviously reflects a simple fact that more and more interactions
come into an action.

Although in a case of the bulk the mean-field theory is too crude to give quantitatively results,
qualitatively it works \cite{Lines}. Moreover, it was shown recently that the mean-field theory adequately
describes the diffuse scattering in MnO in paramagnetic phase \cite{Hohlwein}.

\subsection{Temperature of the magnetic transition}

In the large channels confined nanoparticles are expected to behave as constrained 3D-systems. However with
the channel diameter decreasing one expects a crossover to one-dimensional behavior. In this case the
magnetic fluctuations should destroy the long-range magnetic order and $T_N$ should go to zero. However, in
our case we see that $T_N$ does not extrapolate to zero with channel diameter decreasing (figure
\ref{exponent}a).

Generally the boundary effects decreases the transition temperature. Indeed, one should expect, that in the
vicinity of a nanoparticle surface, the absolute values of the exchange constants decrease due to disorder.
In the phenomenological theory of a finite crystal this effect is described by the term of a positive surface
energy in the thermodynamic potential \cite{Kaganov,Mills}. The disorder is accompanied by the decrease of
the mean moment and the transition temperature at the surface that leads to the decrease of the transition
temperature of the entire nanoparticle.

Another factor decreasing of the transition temperature is a common effect expected for any nanostructured
materials resulted from the limiting the correlation length by the nanoparticle size. It was shown
experimentally for the magnetic nanowires of Ni \cite{Sun}, thin layers of CoO\cite{Ambrose} and numerical
calculation \cite{MacFarland}.

In our case the nanoparticles embedded in the channels of silica matrices with diameters 45-97 {\AA} are
nanoribbons, while nanoparticles within the channels with diameters 24-35 {\AA} are nanowires. In spite of
this topological difference $T_N$ appears above the bulk value for all samples. Moreover, the nanoparticles
of MnO embedded in a porous silica glass with an isotropic shape show the enhanced $T_N$ too\cite{MnO-PRL}.
Obviously that topology of the nanoparticle does not play dominant role.

Neutron diffraction experiments carried out on epitaxial thin films of MnO grown on different substrates
showed that depending on the substrate the N\'eel temperature may be strongly enhanced above that of the bulk
as well as depressed, however it does not depend on the film thickness \cite{Neubeck}. It looks that $T_N$
depends on the non-stoichiometry or/and distortion of the surface between MnO and substrate. Indeed, a system
inhomogeneity can enhance the N\'eel temperature.

\subsection{ Interaction of ferromagnetic moment with antiferromagnetic and structural order parameters}

Because of small dimensions the translation symmetry is violated throughout a significant part of the
nanoparticle volume. The distorted crystal lattice of such objects can be considered as a set of static
translation modes with wave vectors in the interval $(1/{a} - 1/{L})$, where \textit{L} is the characteristic
size of the nanoparticle. In this case new interactions of the order parameter with other magnetic degrees of
freedom, forbidden by the symmetry of the bulk, start to play a role. The similar situation, when the
structure distortions and the "broken exchange bonds" at the nanoparticle surface result in the appearance of
a large number of sub-lattices within nanoparticle was modelled for NiO nanoparticles \cite{Kodama}.

Let us consider the antiferromagnetic phase transition in a lattice with "condensed" structural distortions,
which are described by the order parameters $\eta_{k_s}$ with the corresponding wave vectors $\textbf{k}_s$
(\textit{s} = 1, 2, ...). Antiferromagnetic ordering in this system can be described by an antiferromagnetic
critical order parameters $l_{k}$ with the wave vector \textbf{k}, which has a symmetry of the undistorted
lattice and non-critical, secondary parameters ${l_{k^{'}}^{'}}$, forbidden by the symmetry of the
undistorted lattice.

Generally, the contribution to the thermodynamic potential due to the interaction between the structure
parameters $\eta_{k_s}$, critical parameter $l_k$ and non-critical parameters ${l_{k^{'}}^{'}}$ can be
described by the bilinear invariant with respect to time inversion, namely, $\eta_{k_s} {l_k}{l_{k^{'}}^{'}}$
with ${\textbf{k}_s}+{\textbf{k}}+\textbf{k}^{'}=0$.

In the distorted lattice there are no restrictions for ${\textbf{k}^{'}}$. Therefore the secondary order
parameter with ${\textbf{k}^{'}}= 0$ can participate in the phase transition, i.e. the transition will
accompany by the appearance of the ferromagnetic moment $M$. In the considered model of the distorted lattice
a ferromagnetic moment can be associated with the surface magnetism as well as with the the uncompensated
magnetic moments of different sublattices, as had been proposed by L. N\'eel \cite{Neel}.

The ternary interaction of the structure parameters with the antiferromagnetic order parameter and the
ferromagnetic moment can increase the phase transition temperature that can be readily demonstrated in the
framework of Landau theory. In the first approximation we can consider the simplest thermodynamic potential:

\begin{equation}
\Phi=\Phi_{0}+{\frac{1}{2}}\tau{{l_k}^2}+{\frac{1}{4}}b{{l_k}^4}+f{\eta_k}{l_k}{M}+{\frac{1}{2}}A{M^2},
\label{eq1}
\end{equation}

\noindent Here $b>0$, $A>0$ and \emph{f} are some coefficients, $\tau=(T-T_{bulk})/{T_{bulk}}$. The free
energy (\ref{eq1}) can be minimized with respect to $l_k$ and \textit{M}. Since at the antiferromagnetic
transition ${l_k}{ } = 0 $ we obtain for the temperature of the antiferromagnetic transition::

\begin{equation}
T_N=T_{bulk}(1+{\frac{{(f{\eta_k})}^2}{A}}), \label{eq2}
\end{equation}

Also, taking into consideration new structural parameters (new degrees of freedom) appearing in the distorted
lattice and their interaction with magnetic order parameters one can get the enhanced temperature of the
antiferromagnetic transition with respect to the bulk.

The condition ${\textbf{k}_s}+{\textbf{k}}= 0$ means that the magnetic order induces a translation mode with
the same wave vector.  Unlikely that this type of distortion can be experimentally visible in diffraction at
the small nanoparticle with coherence length limited by the some tens of {\AA}. It is worth to note, the
recent inelastic neutron scattering experiments in the bulk MnO confirmed an instability of the phonon
spectra induced by the magnetic order \cite{Chung}.

In fact there are some competitive factors which influence the transition temperature. The surface
disordering and a limitation of the correlation length to the nanoparticle size lead to a decrease of $T_N$,
while the above mechanism of the ternary interactions due to inhomogeneity leads to an increase of $T_N$. The
experimentally observed increasing in transition temperature with the channel diameter shows that in the
region of the studied matrices the ternary interactions dominate. Generally with an additional increase of
the channel diameter the temperature of the magnetic transition should decrease due to increase of the
correlation length approaching the temperature of the bulk. However, it is necessary  take into consideration
that unlikely the ribbons shape of confined nanoparticles existing in the large channels will still keep at
further channel diameter increasing.

It should be emphasized that in the framework of Landau theory the ferromagnetic moment and the enhanced
N\'eel temperature are related. It is consistent with the observation of the small coercive fields in the
same samples \cite{ESR}. Moreover the ferromagnetic moment at the surface of the nanoparticle in spite of its
antiferromagnetic core, should lead to superparamagnetic behavior of the ensemble of confined nanoparticles
below the transition.

\section{Conclusion}

Neutron diffraction studies of antiferromagnetic MnO confined within MCM-41 type channel matrices with
channel diameters 24 - 87 {\AA} show a magnetic behavior strongly different from that of the bulk. While the
magnetic structure in confinement remains the same as in the bulk, the ordered magnetic moment is noticeably
smaller and the magnetic phase transition is continuous with the increased N\'eel temperature $T_N$ in
contrast to the discontinuous first order transition in the bulk.

With decreasing the channel diameter the critical exponent in the temperature dependence of the magnetic
moment linearly decreases. We attribute the observed transformation of the magnetic transition to the
increasing of anisotropy and the changing of the dimensionality of the magnetic system.

The transformation of the magnetic transition is accompanied by the decrease of $T_N$ with decreasing the
channel diameter. However, for all studied samples $T_N$ remains higher than that for the bulk.

The magnetic disordering and violation of the translation symmetry in the nanoparticle can results in a small
ferromagnetic moment while the dominant magnetic order remains antiferromagnetic. Taking into account the
ternary interaction of the ferromagnetic, antiferromagnetic and associated structural order parameters the
increased N\'eel temperature can be readily explained within the framework of Landau theory.

\begin{acknowledgements}
The authors thank C. Alba-Simionesco, N. Brodie and G. Dosseh who prepared and characterized MCM and SBA
matrices. They are very grateful to Prof. S. B. Vakhrushev for critical reading of the manuscript and for
fruitful discussions. The work was supported by the RFBR (Grants 04-02-16550 and SS-1671-2003.2) and the
INTAS (Grant No. 2001-0826).
\end{acknowledgements}

\end{document}